\documentclass[twocolumn,showpacs,preprintnumbers,amsmath,amssymb,nofootinbib]{revtex4-2}
\usepackage{comment}
\usepackage{graphicx}
\usepackage{bm}
\usepackage{xcolor}
\usepackage{hyperref}
\hypersetup{colorlinks=true,linkcolor=blue,urlcolor=blue,citecolor=blue}

\newcommand{\be}{\begin{equation}}
\newcommand{\ee}{\end{equation}}

\newcommand{\Fib}{\mathrm{Fib}}
\newcommand{\GE}{G_{E_8}}
\newcommand{\kk}{\widehat{k}}

\newcommand{\AFib}{A_{\Fib}}
\newcommand{\one}{\mathbb{I}}

\begin{document}
\title{UV completion of 2D Ising CFT:\\ a golden $E_8$ massless $S$-matrix}

\newcommand{\sharedaffil}{\vspace{3pt}\\${}^{\dagger}$Department of Physics, Ewha Womans University, Seoul 03760, Korea\\
${}^{\sharp}$RIBS \& Department of Physics and Astronomy, Seoul National University, Seoul 08826, Korea\\
${}^{\flat}$School of Science, Huzhou Normal University, Huzhou 313000, China}

\author{Changrim Ahn\,$^{\dagger}$}
\affiliation{\sharedaffil}
\author{Minkyoo Kim\,$^{\sharp,\flat}$}
\affiliation{\sharedaffil}

\begin{abstract}
\centering\begin{minipage}{\dimexpr\paperwidth-6.0cm}
\noindent
We present a full classification of UV complete QFTs that RG flow to the 2D Ising CFT by solving the bootstrap equations for massless right--left $S$-matrices. For the Ising model with $E_8$ spectrum we find exactly four completions, arising from higher-$T\bar T$-type deformations, including a previously unknown ``golden flow'' whose UV fixed point is a diagonal $su(2)$ coset CFT ($c=25/14$) along $\Delta_{\rm rel}=2/7$. A universal Fibonacci/$E_8$ structure governs the R--L adjacency matrices and the $Y$-system periods, so that the $E_8$ symmetry persists across all RG scales.
\end{minipage}
\end{abstract}
\maketitle
\vspace{-5pt}

\section{Introduction}

The two-dimensional (2D) Ising conformal field theory (CFT) is one of the most fundamental building blocks of string theories as well as many other applications. Our prime goal is to find a complete list of UV CFTs and deformations which RG flow into the 2D Ising CFT in terms of concrete $S$-matrices based on the integrability. For this purpose, we need to find all exact $S$-matrices which can describe the Ising CFT in the IR limit and irrelevant deformations. 

A systematic approach was proposed in ~\cite{AhnLeClair2022} where the RG flows are constructed by $S$-matrices between massless particles describing the IR theory. 
In the massless limit, the excitations split into right (R) and left (L) movers.
The right--right (R--R) and left--left (L--L) $S$-matrix amplitudes are the same as 
those of the massively deformed IR CFT. 
Deforming an integrable field theory by the irrelevant fields built from the higher
$T\bar T$-type conserved charges generates CDD-like scalar factors in the 
massive $S$-matrices ~\cite{SmirnovZam,CNST}.
These factors contribute the right--left (R--L) and left--right (L--R) $S$-matrices in the massless limit ~\cite{AhnLeClair2022}.
Thermodynamic Bethe ansatz (TBA) equations derived from these amplitudes identify UV complete theories connected to the IR CFT by the RG flows.

The 2D Ising CFT can maintain integrability upon two particular deformations.
Deformed by the energy operator, this is described by a massive free Majorana fermion. Using this simple $S$-matrix, one can still compute various interesting correlation functions. 
The other deformation is obtained by the magnetic spin operator.
This integrable field theory is given by the Zamolodchikov's celebrated $S$-matrix with eight self-conjugate particles, whose masses are given by the Perron-Frobenius eigenvector of the $E_8$ symmetry~\cite{Zam1989}.
Some of these excitations have been experimentally observed with a golden mass ratio $m_2/m_1=2\cos(\pi/5)=\varphi$~\cite{Coldea2010}.

The Ising CFT is thus
described by \emph{two} distinct massless theories, the free-fermion one
and the $E_8$ one, recovered as the massless (conformal) limits of the
two massive deformations. For the free-fermion one with the (R--R) and (L--L) $S$-matrices $-1$, the R--L $S$-matrices that generate RG flows into the tricritical Ising CFT with $c=7/10$ ~\cite{alyosha} or N=1 supersymmetric CFTs with $c=3/2$ ~\cite{akrz,andre} have been previously found. 

Our main interest is to represent the Isng CFT as a scattering theory of the eight particles in the massless limit $m_1\to 0$ while maintaining the mass ratios $m_a/m_1$, which are proportional to the $E_8$ Perron-Frobenius eigenvector elements.
While the (R--R) and (L--L) amplitudes are the same as Zamolodchikov's $S$-matrix, explicit expressions of the (R--L) and (L--R) $S$-matrices have been missing. 
Although a previous study found three UV
completions---diagonal, minimal, and saturated---without explicit $S$-matrices, complete analysis of the RG flows is not available~\cite{AhnLeClair2022}. For the minimal case, a conjectured TBA proposed in ~\cite{ravanini} remains unproven. 
Our result in this letter provides not only explicit $S$-matrices and TBAs for these three cases but also discovers one additional UV completion, the ``golden''. We also find a universal relation based on the $E_8$ symmetry that connects all these four.

For a full classification, we need to find complete solutions of the massless bootstrap equations which have been used in $A_n$ scattering theories ~\cite{Ahn2025}.
As a summary, we (i) construct the finite positive R--L
amplitudes satisfying the bootstrap and saturation bound which generate not only the three UV completions but also the new ``golden'' solution; (ii) derive
its TBA/$Y$-system by inversion relations and read off the UV central charge and
the relevant deforming field; (iii) identify the golden UV candidate
as the diagonal coset $(su(2)_5\times su(2)_5)/su(2)_{10}$ from this TBA
data and its $\Delta=2/7$ operator content, checked
against the genuine affine-branching spectrum; and (iv) check the relevant massless $Y$-system periods by exact integer/tropical $Y$-pattern computations.

\begin{table*}[t!]
\centering
\caption{Complete R--L amplitude tables of the minimal and golden completions
($a\le b$, $S_{ba}=S_{ab}$), $S_{ab}=\prod_p[p]^{\,n_{ab}(p)}$ with the shorthand $[p]\equiv F_p(\theta)$ for compactness. The exponents define the matrices $n(p)$ used below.}
\label{tab:RL}
\renewcommand{\arraystretch}{1.12}
{\scriptsize
\begin{tabular}{ |l|l|l||l|l|l| }
\hline\hline
 & minimal & golden & & minimal & golden\\
\hline
$S_{11}$ & $[1][11]$ & $[7][13]$ & $S_{36}$ & $[5][7]^2[9][11][13]^2[15]$ & $[1][3][5][7]^2[9]^2[11]^3[13]^3[15]$\\
$S_{12}$ & $[7][13]$ & $[1][7][11][13]$ & $S_{37}$ & $[3][5][7][9]^2[11]^2[13]^2[15]$ & $[3][5]^2[7]^2[9]^3[11]^3[13]^3[15]^2$\\
$S_{13}$ & $[2][10][12]$ & $[6][8][12][14]$ & $S_{38}$ & $[4][6]^2[8]^2[10]^2[12]^2[14]^3$ & $[2][4]^2[6]^2[8]^3[10]^4[12]^4[14]^4$\\
$S_{14}$ & $[6][10][14]$ & $[4][8][10][12][14]$ & $S_{44}$ & $[1][5][7][9][11]^2[13][15]$ & $[3][5][7]^2[9]^2[11]^2[13]^3[15]$\\
$S_{15}$ & $[3][9][11][13]$ & $[5][7][9][11][13][15]$ & $S_{45}$ & $[4][6][8]^2[10][12]^2[14]^2$ & $[2][4][6]^2[8]^2[10]^3[12]^3[14]^3$\\
$S_{16}$ & $[6][8][12][14]$ & $[2][6][8][10][12]^2[14]$ & $S_{46}$ & $[3][5][7][9]^2[11]^2[13]^2[15]$ & $[3][5]^2[7]^2[9]^3[11]^3[13]^3[15]^2$\\
$S_{17}$ & $[4][8][10][12][14]$ & $[4][6][8][10]^2[12][14]^2$ & $S_{47}$ & $[3][5][7]^2[9]^2[11]^2[13]^3[15]$ & $[1][3][5]^2[7]^3[9]^3[11]^4[13]^4[15]^2$\\
$S_{18}$ & $[5][7][9][11][13][15]$ & $[3][5][7][9]^2[11]^2[13]^2[15]$ & $S_{48}$ & $[2][4][6]^2[8]^2[10]^3[12]^3[14]^3$ & $[2][4]^2[6]^3[8]^4[10]^4[12]^5[14]^5$\\
$S_{22}$ & $[1][7][11][13]$ & $[1][7]^2[11][13]^2$ & $S_{55}$ & $[1][3][5][7][9]^2[11]^3[13]^2[15]$ & $[3][5]^2[7]^3[9]^3[11]^3[13]^4[15]^2$\\
$S_{23}$ & $[6][8][12][14]$ & $[2][6][8][10][12]^2[14]$ & $S_{56}$ & $[4][6]^2[8]^2[10]^2[12]^2[14]^3$ & $[2][4]^2[6]^2[8]^3[10]^4[12]^4[14]^4$\\
$S_{24}$ & $[4][8][10][12][14]$ & $[4][6][8][10]^2[12][14]^2$ & $S_{57}$ & $[2][4][6]^2[8]^2[10]^3[12]^3[14]^3$ & $[2][4]^2[6]^3[8]^4[10]^4[12]^5[14]^5$\\
$S_{25}$ & $[5][7][9][11][13][15]$ & $[3][5][7][9]^2[11]^2[13]^2[15]$ & $S_{58}$ & $[3][5]^2[7]^3[9]^3[11]^3[13]^4[15]^2$ & $[1][3]^2[5]^3[7]^4[9]^5[11]^6[13]^6[15]^3$\\
$S_{26}$ & $[2][6][8][10][12]^2[14]$ & $[2][6]^2[8]^2[10][12]^3[14]^2$ & $S_{66}$ & $[1][3][5][7]^2[9]^2[11]^3[13]^3[15]$ & $[1][3][5]^2[7]^4[9]^3[11]^4[13]^5[15]^2$\\
$S_{27}$ & $[4][6][8][10]^2[12][14]^2$ & $[4]^2[6][8]^2[10]^3[12]^2[14]^3$ & $S_{67}$ & $[3][5]^2[7]^2[9]^3[11]^3[13]^3[15]^2$ & $[3]^2[5]^3[7]^3[9]^5[11]^5[13]^5[15]^3$\\
$S_{28}$ & $[3][5][7][9]^2[11]^2[13]^2[15]$ & $[3][5]^2[7]^2[9]^3[11]^3[13]^3[15]^2$ & $S_{68}$ & $[2][4]^2[6]^2[8]^3[10]^4[12]^4[14]^4$ & $[2][4]^3[6]^4[8]^5[10]^6[12]^6[14]^7$\\
$S_{33}$ & $[1][3][9][11]^2[13]$ & $[5][7]^2[9][11][13]^2[15]$ & $S_{77}$ & $[1][3][5]^2[7]^3[9]^3[11]^4[13]^4[15]^2$ & $[1][3]^2[5]^3[7]^5[9]^5[11]^6[13]^7[15]^3$\\
$S_{34}$ & $[5][7][9][11][13][15]$ & $[3][5][7][9]^2[11]^2[13]^2[15]$ & $S_{78}$ & $[2][4]^2[6]^3[8]^4[10]^4[12]^5[14]^5$ & $[2]^2[4]^3[6]^5[8]^6[10]^7[12]^8[14]^8$\\
$S_{35}$ & $[2][4][8][10]^2[12]^2[14]$ & $[4][6]^2[8]^2[10]^2[12]^2[14]^3$ & $S_{88}$ & $[1][3]^2[5]^3[7]^4[9]^5[11]^6[13]^6[15]^3$ & $[1][3]^3[5]^5[7]^7[9]^8[11]^9[13]^{10}[15]^5$\\
\hline\hline
\end{tabular}}
\end{table*}

\section{New solutions of the bootstrap}
The elementary CDD block can be expressed with
\be
F_p(\theta)\equiv\frac{\sinh\theta-i\sin(\pi p/30)}{\sinh\theta+i\sin(\pi p/30)},
\qquad p\in\mathbb{Z}.
\label{eq:Fp}
\ee
Each block is crossing-unitary, $F_p(\theta)F_p(\theta+i\pi)=1$, and pole-free in the physical strip. 
Using $F_{30-p}=F_p$, $F_{-p}=F_p^{-1}$, $F_0=F_{30}=1$, $F_{p+60}=F_p$,  
we can express every $S$-matrix element as a product of each $F_p$ ($p=1,\cdots,15)$ with non-negative integer power $n_{ab}(p)$, $S_{ab}=\prod_{p=1}^{15} F_p^{\,n_{ab}(p)}$.
Therefore, the (R--L) scatterings $S_{ab}$ are essentially encoded in these exponents $n_{ab}(p)$.  We find all possible solutions $n_{ab}$ from the bootstrap equations along the following three steps.

\emph{(1) Fix $S_{11}$.} The self-fusion $A_1A_1\to A_1$ gives the first 
bootstrap equation
\begin{equation}
  S_{11}(\theta)=S_{11}\left(\theta+\frac{i\pi}{3}\right)S_{11}\left(\theta-\frac{i\pi}{3}\right).
  \label{eq:shift}
\end{equation}
Each solution of this provides a seed to construct all other amplitudes.
The saturation bound below limits the seeds to at most four $F_p$ blocks. 
The new solutions already appear at two blocks: after full directed propagation,
pole-free positivity, and the saturation bound, the surviving two-block seeds
are the minimal $F_1F_{11}$ and the new golden $F_7F_{13}$, while the saturated
solution is the entrywise square $(F_1F_{11})^2$. The diagonal completion
requires no search: its R--L amplitudes are the inverses of the massive $E_8$
amplitudes,
$S^{\rm diag}_{ab}=\big(S^{E_8}_{ab}\big)^{-1}$, which are pole-free in the
physical strip and solve the same fusing equations, giving the seed
$F_2F_{10}F_{12}$. 

\emph{(2) Generate all amplitudes from $S_{11}$.} Each heavier particle is a
bound state reached by an equal-mass fusion $A_aA_a\to A_c$, realized as a
shift with the fusion angle $\overline{u}^{a}_{ac}$;
$A_1A_1\to A_{2,3}$ with $m=6,1$; $A_2A_2\to A_{4,5,6}$ with $m=7,4,1$;
$A_3A_3\to A_7$ with $m=2$; and $A_4A_4\to A_8$ with $m=1$ (the complete
fusion angle table is in the SM~\cite{SM}). 
The complete minimal and golden tables obtained from this procedure are listed
in TABLE~\ref{tab:RL}; the saturated table is the entrywise square of the
minimal one.

\emph{(3) Consistency} There remain many bootstrap equations 
\be
S_{dc}(\theta)=S_{da}\!\big(\theta+i\overline{u}^b_{ac}\big)\,
              S_{db}\!\big(\theta-i\overline{u}^a_{bc}\big),
\label{eq:boot}
\ee
with the generic fusion angles $\overline{u}^b_{ac}$. These over-determined equations provide a strict consistency check: all
$1792$ directed fusing identities hold as exact integer-vector identities for
each completion in Table~\ref{tab:RL}.
The resulting S-matrices are written as
\begin{equation}
S_{ab}(\theta)=\prod_{p=1}^{15} F_p(\theta)^{\,n_{ab}(p)},
\label{bsm}
\end{equation}
with integer block exponents \(n_{ab}(p)\) specifying the multiplicity of each elementary CDD block \(F_p\).	

UV completeness requires the integrated R--L kernel $\kk_{ab}=\sum_p n_{ab}(p)$
to satisfy $\kk_{ab}\le(2M)_{ab}$, with $M=(2\one-\GE)^{-1}$, the inverse $E_8$ 
Cartan matrix where $\GE$ is the mass-ordered adjacency matrix ~\cite{AhnLeClair2022}. 
Seeds violating the saturation bound, $\kk_{11}>4$, produce a Hagedorn divergence in the UV limit.
A simple check over all possible $S_{11}$ against the saturation bounds  leaves exactly four pole-free UV completions, with
$c_{\rm UV}=21/22,\,25/14,\,15/2,\,31/2$ for the minimal, golden,
diagonal, and saturated seeds (SM).

\section{TBA and the $Y$-system}

\subsection{Inversion relations and the universal kernel}

In the standard derivation of the TBA equations, the pseudo-energies of the R-particles couple to those of the L-particles. 
The Fourier transform of the R--L kernel matrix, whose elements are the logarithmic derivatives of the $S_{ab}$, can be obtained by \eqref{bsm}
\begin{equation}
\Psi(q)=\sum_{p=1}^{15} n(p)\left(q^{15-p}+q^{p-15}\right), \quad
q=e^{\pi\omega/30}
\label{eq:PsiBlock}
\end{equation}
where each factor \(q^{15-p}+q^{p-15}\) comes from the Fourier transform of $F_p(\theta)$.

The $E_8$ symmetry appears in universal TBA equations where kernels connecting the R and L pseudo-energies are given by \(\big[(q+q^{-1})\one-\GE\big]\,\Psi(q)\).
This can be rewritten with Chebyshev polynomials $T_n$
\be
2\sum_{p=1}^{15} n(p)(2T_1(x)\one-\GE)T_{15-p}(x),\quad x=\frac{q+q^{-1}}{2}
\label{eq:e8inversion}
\ee
where $\GE$ is the adjacency matrix of the $E_8$.
It turns out that the exponent matrices $n(p)$ given in TABLE~\ref{tab:RL} satisfy interesting recursion relations
\be
(1+\delta_{p,14})n(p+1)+n(p-1)=\GE n(p),
\ee
where \(p=1,\cdots,15\) with \(n(0)=n(16)\equiv 0\) are implied.
These relations and Chebyshev identity \(2T_1(x)T_n(x)=T_{n+1}(x)+T_{|n-1|}(x)\)
make the intermediate Chebyshev coefficients all cancel, leaving only the boundary term, 
\be
\big[(q+q^{-1})\one-\GE\big] \Psi(q)
=(q^{15}+q^{-15})\mathbb{N},
\label{eq:univ}
\ee
where we use $\mathbb{N}=n(1)$ for simplicity. 
For the minimal $S$-matrix in TABLE~\ref{tab:RL}, the $[1]$ factor appears only in $S_{aa}$, hence, $\mathbb{N}=n(1)=\one_8$. For the golden solution, the exponent matrix can be written as
\be
\mathbb{N}=\left(
\begin{smallmatrix}
0&1&0&0&0&0&0&0\\
1&1&0&0&0&0&0&0\\
0&0&0&0&0&1&0&0\\
0&0&0&0&0&0&1&0\\
0&0&0&0&0&0&0&1\\
0&0&1&0&0&1&0&0\\
0&0&0&1&0&0&1&0\\
0&0&0&0&1&0&0&1
\end{smallmatrix}
\right)=\AFib^{\oplus4},\quad \AFib=\begin{pmatrix}0&1\\1&1\end{pmatrix}.
\ee
We call this structure golden/Fibonacci because the golden kernel obeys
\be
\mathbb{N}^2=\mathbb{N}+\one,
\ee
where the four pairs, connected by $\mathbb{N}$ 
\[
\{1,2\},\qquad \{3,6\},\qquad \{4,7\},\qquad \{5,8\},
\]
are precisely the golden doublets of the mass spectrum,
\begin{equation}
\frac{m_2}{m_1}=\frac{m_6}{m_3}=\frac{m_7}{m_4}=\frac{m_8}{m_5}=\varphi . \nonumber
\end{equation}
Thus the golden ratio visible in the massive \(E_8\) spectrum reappears here as
a kernel-level Fibonacci structure.

\begin{figure}[t]
\centering
\includegraphics[width=1\columnwidth]{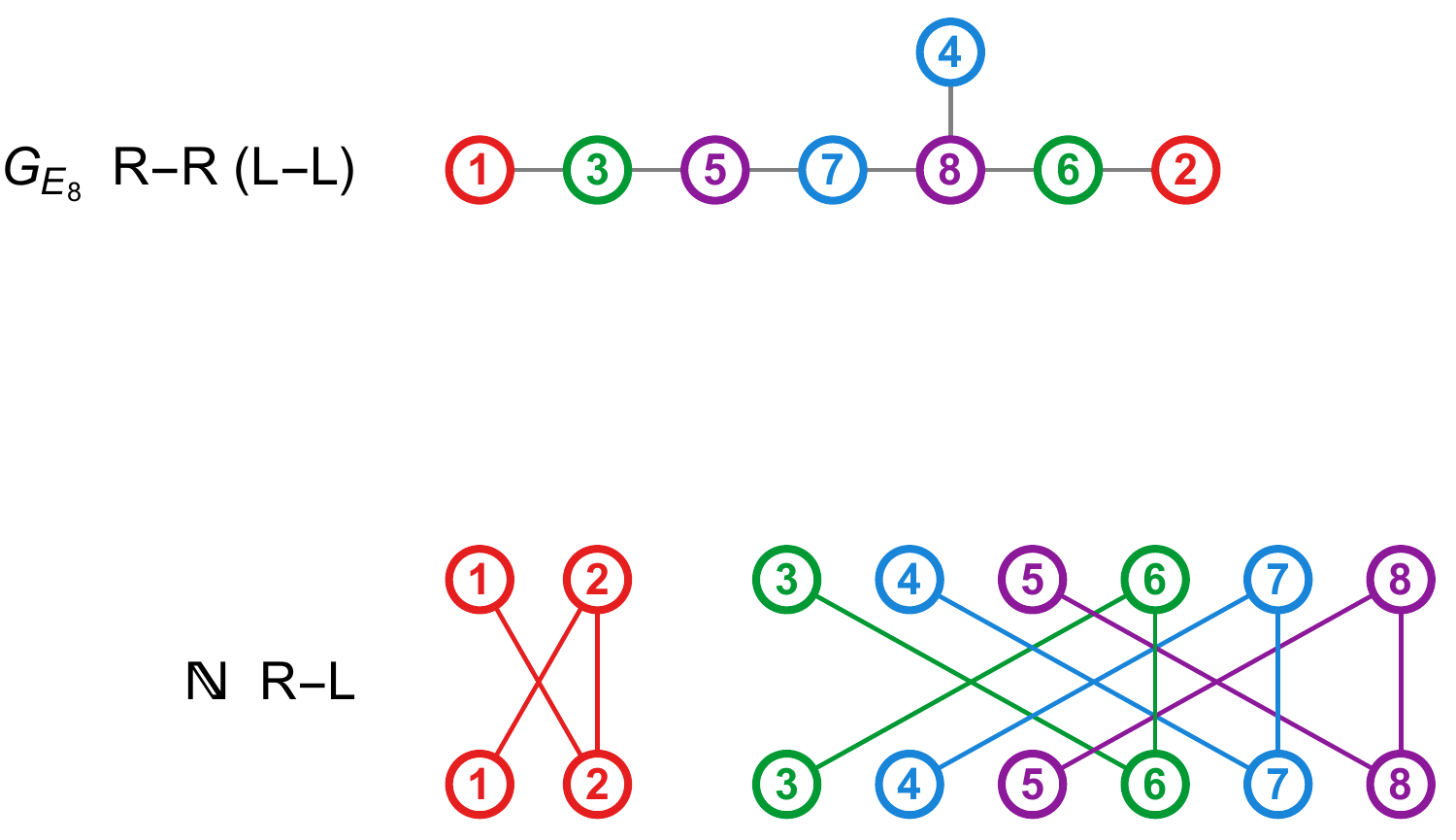}
\caption{Top: the mass-ordered \(E_8\) adjacency graph, with node labels denoting the
particle mass ordering. Bottom: the golden R--L coupling graph in the same
mass-ordered basis. The colored links indicate the four Fibonacci blocks
associated with the golden doublets.}
\label{fig:fib}
\end{figure}

The $Y$-systems for each UV-complete solution can be derived as follows:
\be
Y^R_a(\theta+\frac{i\pi}{30})Y^R_a(\theta-\frac{i\pi}{30})=\prod_{b=1}^{8}
\frac{(1+Y^R_b)^{(G_{E_8})_{ab}}}{(1+1/Y^L_b)^{\mathbb{N}_{ab}}},
\label{ysystem}
\ee
where the adjacency graphs connecting the R- and L-nodes are plotted in FIG.\ref{fig:fib}. 
It is remarkable that all the \(\mathbb{N}\) adjacency matrices for the UV complete theories can be written as a universal form 
\be
\mathbb{N}^{[n]}=2T_n(\GE/2),
\qquad n=10,6,1,0.
\label{eq:chebfold}
\ee
Furthermore, only the above $T_n(\GE/2)$ have non-negative matrix elements among all positive divisors \(n\) of \(30\). 
This is consistent with the UV completeness since the $Y$-systems become singular if some $\mathbb{N}_{ab}$ become negative.

\subsection{UV central charge and relevant field from exact periodicity}
\label{sec:periodicity}
In the UV plateau the constant TBA solution gives, through the Rogers-dilogarithm
sum, $c_{\rm UV}=25/14$ for the golden flow. The dimension of the UV relevant field can be read directly from the periodicity of the $Y$-system \eqref{ysystem}.
For the golden completion the exact period analysis gives
$P_{\rm TBA}=35$, and therefore
\be
\Delta_{\rm rel}=2-\frac{60}{P_{\rm TBA}}=\frac{2}{7}.
\label{eq:Drel}
\ee

The same analysis gives \(\Delta_{\rm rel}=2/11\) and \(1\) for the minimal and
diagonal completions (\(P_{\rm TBA}=33,60\)), while the saturated case lies
outside the finite ADE periodic family and is treated as the
\(P_{\rm TBA}=\infty\) marginal endpoint (\(\Delta_{\rm rel}=2\)).
The four completions are collected in TABLE~\ref{tab:completions}.

\begin{table}[h]
\centering
\begin{tabular}{lcccc}
\hline\hline
completion & $\mathbb{N}$ & $n$ & $c_{\rm UV}$ & $\Delta_{\rm rel}$\\
\hline
minimal & $\one$ & $10$ & $21/22$ & $2/11$\\
golden & $\AFib^{\oplus4}$ & $6$ & $25/14$ & $2/7$\\
diagonal & $\GE$ & $1$ & $15/2$ & $1$\\
saturated & $2\one$ & $0$ & $31/2$ & $2$\\
\hline\hline
\end{tabular}
\caption{The four UV completions: constant kernel $\mathbb{N}=2T_n(\GE/2)$, UV central
charge, and relevant dimension $\Delta_{\rm rel}=2-60/P_{\rm TBA}$ with
$P_{\rm TBA}=33,35,60,\infty$. The saturated entry is marginal.}
\label{tab:completions}
\end{table}

\subsection{Numerical golden flow}
\begin{figure}[t]
\centering
\includegraphics[width=0.95\columnwidth]{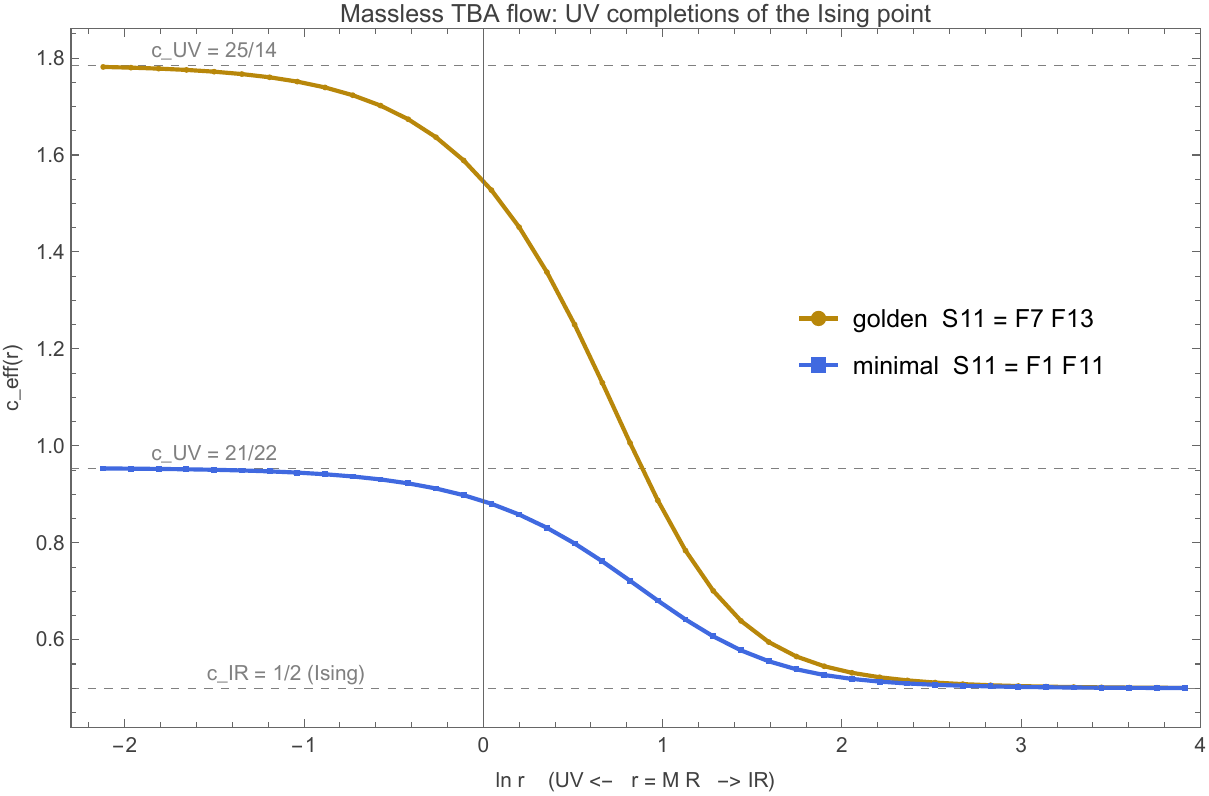}
\caption{Effective central charge $c_{\rm eff}(r)$ from the numerical TBA.  The
minimal curve is included as a reference, while the golden flow generated by
$F_7F_{13}$ rises from the infrared Ising value $1/2$ to the new ultraviolet
plateau $25/14$.}
\label{fig:ceff}
\end{figure}

Solving the golden TBA numerically, we find that the effective central charge rises 
monotonically from $c_{\rm IR}=1/2$ to the UV plateau $25/14$ for the golden and $21/22$ for the minimal flows (Fig.~\ref{fig:ceff}).  
The golden $L$-functions are bell-shaped and mass-ordered (SM, Fig.~S1), as expected for a relevant, minimal-type massless flow.

\subsection{Identification of the golden UV CFT}
The UV CFT for the minimal flow is well-known to be a coset CFT
with the relevant coset field
\be
\frac{E_{8,1}\times E_{8,2}}{E_{8,3}}+\lambda\Phi_{\rm rel},\quad
\Phi_{\rm rel}=\left[
\frac{(1;\circ)\otimes (2;\circ)}{(3;{\rm Adj} ) } \right],
\ee
where \((k;\circ)\) denotes a singlet representation of the $E_8$ WZW CFT with a level $k$.

From the UV conformal data for the golden flow
\be
(c_{\rm UV},\Delta_{\rm rel})=\left(\tfrac{25}{14},\tfrac{2}{7}\right),
\label{eq:uvfingerprint}
\ee
we identify it as another coset CFT
\be
\frac{su(2)_5\times su(2)_5}{su(2)_{10}},\qquad c=\frac{25}{14}.
\label{eq:ccoset}
\ee

The central charge alone is not claimed to determine the UV theory uniquely.
We therefore check whether the relevant field with $\Delta=2/7$ can exist in 
the field content of the coset CFT. 
In terms of the $su(2)$ representations, the highest weight states of the coset can be written as
$(\ell_1,\ell_2;\ell_3)$ with $0\le\ell_{1,2}\le5$, $0\le\ell_3\le10$,
$\ell_1+\ell_2+\ell_3\in2\mathbb{Z}$ and the field identification
\be
(\ell_1,\ell_2;\ell_3)\sim(5-\ell_1,5-\ell_2;10-\ell_3),
\label{eq:fieldid}
\ee
the conformal weight (mod $1$) is
\be
h=\frac{\ell_1(\ell_1+2)}{28}+\frac{\ell_2(\ell_2+2)}{28}
 -\frac{\ell_3(\ell_3+2)}{48}.
\label{eq:hcoset}
\ee
We find that
\be
(1,3;4)\sim(3,1;4)\sim(2,4;6)\sim(4,2;6)
\label{eq:phi-orbit-doubled}
\ee
give the primary field with $\Delta=2h=2/7$. 

Another interesting point is known in mathematical literature that the period of the coupled $Y$-systems is given by sum of the individual periods (see ~\cite{GalashinPylyavskyy} and references therein).
Accordingly, the period of the golden $Y$-system can be interpreted as $P_{\rm TBA}=5+30=h_{A_4}+h_{E_8}$, a sum of two dual Coxeter numbers.
Along with a level-rank duality between $su(2)_5$ and $su(5)_2$, the period and the UV diagonal $su(2)_5$ coset are consistent.

\section{Discussion}
The four sets of the R--L $S$-matrices generate the UV-complete integrable massless flows from the Ising CFT deformed by the higher-$T\bar T$-type deformations.
The new golden flow, which had not even been conjectured before, moves toward the UV CFT along a relevant $\Delta_{\rm rel}=2/7$ field direction.
The corresponding UV CFT is identified with a diagonal $su(2)_5$ coset which flows from \(c=25/14\) to \(c=1/2\).
The golden ratio visible in the \(E_8\) mass spectrum reappears in
the UV completion as a kernel-level Fibonacci structure,
\(N_{\rm Fib}=A_{\rm Fib}^{\oplus4}\) and \(N_{\rm Fib}^2=N_{\rm Fib}+I\).

We have also found that the four UV complete RG flows have an interesting common feature, namely, all the R--L adjacency matrices are given by \eqref{eq:chebfold}
along with the TBA periods given by a universal formula
\(P_{\rm TBA}=30+30/n\) for \(n=1,6,10\), with the saturated case \(n=0\)
understood as the marginal \(P_{\rm TBA}=\infty\) endpoint.

In conclusion, the beautiful $E_8$ structure discovered in the 2D Ising CFT with the magnetic deformation~\cite{Zam1989} persists even in the UV completion generated by the same theory deformed by higher irrelevant deformations through the UV-complete massless flows constructed here.

\begin{acknowledgments}
This work is supported by the National
Research Foundation of Korea (NRF) through grants RS-2026-25472596 (CA) and RS-2025-25414114 (MK).
\end{acknowledgments}
\vspace{5pt}
\noindent\textbf{Corresponding Authors:}\\
${}^{\dagger}$\href{mailto:ahn@ewha.ac.kr}{ahn@ewha.ac.kr},\,
${}^{\sharp}$\href{mailto:minkyookim@snu.ac.kr}{minkyookim@snu.ac.kr}.

\clearpage

\onecolumngrid
\pagenumbering{roman}
\setcounter{page}{1}
\setcounter{section}{0}
\setcounter{equation}{0}
\setcounter{figure}{0}
\setcounter{table}{0}
\makeatletter
\@removefromreset{equation}{section}
\@removefromreset{figure}{section}
\@removefromreset{table}{section}
\makeatother
\renewcommand{\thesection}{S\arabic{section}}
\renewcommand{\theequation}{S\arabic{equation}}
\renewcommand{\thefigure}{S\arabic{figure}}
\renewcommand{\thetable}{S\arabic{table}}

\begin{center}
\vspace{6pt}
\Large\textbf{Supplemental Material (SM)}\\[12pt]
\normalsize\textbf{UV completion of 2D Ising CFT: a golden $E_8$ massless $S$-matrix}\\[8pt]
Changrim Ahn and Minkyoo Kim
\end{center}
\vspace{12pt}

\section{Saturation bounds for $E_8$ bootstraps}
The mass-ordered $E_8$ spectrum, given by the components of the Perron-Frobenius eigenvector, is
\begin{align}
&m_1=1,\quad m_2=2\cos\tfrac{\pi}{5},\quad m_3=2\cos\tfrac{\pi}{30},\nonumber\\
&m_4=4\cos\tfrac{\pi}{5}\cos\tfrac{7\pi}{30},\quad m_5=4\cos\tfrac{\pi}{5}\cos\tfrac{2\pi}{15},\nonumber\\
&m_6=4\cos\tfrac{\pi}{5}\cos\tfrac{\pi}{30},\quad m_7=8\cos^2\tfrac{\pi}{5}\cos\tfrac{7\pi}{30},\nonumber\\
&m_8=8\cos^2\tfrac{\pi}{5}\cos\tfrac{2\pi}{15}.
\end{align}
The mass-ordered adjacency matrix $\GE$ and the inverse Cartan matrix are given by
\be
\GE=\begin{pmatrix}
0&0&1&0&0&0&0&0\\
0&0&0&0&0&1&0&0\\
1&0&0&0&1&0&0&0\\
0&0&0&0&0&0&0&1\\
0&0&1&0&0&0&1&0\\
0&1&0&0&0&0&0&1\\
0&0&0&0&1&0&0&1\\
0&0&0&1&0&1&1&0
\end{pmatrix},
\quad
M=\begin{pmatrix}
2&2&3&3&4&4&5&6\\
2&4&4&5&6&7&8&10\\
3&4&6&6&8&8&10&12\\
3&5&6&8&9&10&12&15\\
4&6&8&9&12&12&15&18\\
4&7&8&10&12&14&16&20\\
5&8&10&12&15&16&20&24\\
6&10&12&15&18&20&24&30
\end{pmatrix}.
\ee

The fusion angles that generate the bootstrap equations are listed in TABLE~\ref{tab:fusion}. 
\begin{table}[!ht]
\centering
{\footnotesize
\setlength{\tabcolsep}{4pt}
\begin{tabular}{c|ccc||c|ccc||c|ccc||c|ccc}
\hline\hline
$\{a,b,c\}$ & $u^c_{ab}$ & $u^b_{ac}$ & $u^a_{bc}$ &
$\{a,b,c\}$ & $u^c_{ab}$ & $u^b_{ac}$ & $u^a_{bc}$ &
$\{a,b,c\}$ & $u^c_{ab}$ & $u^b_{ac}$ & $u^a_{bc}$ &
$\{a,b,c\}$ & $u^c_{ab}$ & $u^b_{ac}$ & $u^a_{bc}$\\
\hline
$\{1,1,1\}$ & 20 & 20 & 20 & $\{1,7,8\}$ & 5 & 26 & 29 & $\{3,3,3\}$ & 20 & 20 & 20 & $\{4,5,5\}$ & 19 & 19 & 22\\
$\{1,1,2\}$ & 12 & 24 & 24 & $\{1,8,8\}$ & 16 & 16 & 28 & $\{3,3,5\}$ & 14 & 23 & 23 & $\{4,5,8\}$ & 9 & 25 & 26\\
$\{1,1,3\}$ & 2 & 29 & 29 & $\{2,2,2\}$ & 20 & 20 & 20 & $\{3,3,6\}$ & 12 & 24 & 24 & $\{4,7,7\}$ & 18 & 18 & 24\\
$\{1,2,2\}$ & 18 & 18 & 24 & $\{2,2,4\}$ & 14 & 23 & 23 & $\{3,3,7\}$ & 4 & 28 & 28 & $\{4,7,8\}$ & 14 & 21 & 25\\
$\{1,2,3\}$ & 14 & 21 & 25 & $\{2,2,5\}$ & 8 & 26 & 26 & $\{3,4,5\}$ & 16 & 21 & 23 & $\{5,5,5\}$ & 20 & 20 & 20\\
$\{1,2,4\}$ & 8 & 25 & 27 & $\{2,2,6\}$ & 2 & 29 & 29 & $\{3,5,7\}$ & 13 & 22 & 25 & $\{5,5,8\}$ & 12 & 24 & 24\\
$\{1,3,4\}$ & 13 & 21 & 26 & $\{2,3,3\}$ & 19 & 19 & 22 & $\{3,5,8\}$ & 5 & 27 & 28 & $\{5,6,7\}$ & 17 & 21 & 22\\
$\{1,3,5\}$ & 3 & 28 & 29 & $\{2,3,6\}$ & 9 & 25 & 26 & $\{3,6,6\}$ & 18 & 18 & 24 & $\{5,8,8\}$ & 18 & 18 & 24\\
$\{1,4,4\}$ & 17 & 17 & 26 & $\{2,4,7\}$ & 5 & 27 & 28 & $\{3,6,8\}$ & 8 & 25 & 27 & $\{6,6,6\}$ & 20 & 20 & 20\\
$\{1,4,5\}$ & 11 & 22 & 27 & $\{2,5,6\}$ & 16 & 19 & 25 & $\{3,8,8\}$ & 17 & 17 & 26 & $\{6,6,8\}$ & 14 & 23 & 23\\
$\{1,4,6\}$ & 7 & 25 & 28 & $\{2,6,7\}$ & 13 & 21 & 26 & $\{4,4,4\}$ & 20 & 20 & 20 & $\{6,7,8\}$ & 16 & 21 & 23\\
$\{1,5,6\}$ & 14 & 19 & 27 & $\{2,6,8\}$ & 3 & 28 & 29 & $\{4,4,6\}$ & 16 & 22 & 22 & $\{7,7,7\}$ & 20 & 20 & 20\\
$\{1,5,7\}$ & 4 & 27 & 29 & $\{2,7,7\}$ & 17 & 17 & 26 & $\{4,4,7\}$ & 12 & 24 & 24 & $\{7,8,8\}$ & 19 & 19 & 22\\
$\{1,6,7\}$ & 9 & 23 & 28 & $\{2,7,8\}$ & 11 & 22 & 27 & $\{4,4,8\}$ & 2 & 29 & 29 & $\{8,8,8\}$ & 20 & 20 & 20\\
\hline\hline
\end{tabular}}
\caption{The $56$ $E_8$ fusion triples and their external angles
$u^c_{ab}$ (units of $\pi/30$), obeying $u^c_{ab}+u^b_{ac}+u^a_{bc}=60$.}
\label{tab:fusion}
\end{table}

All $S_{1a}$ can satisfy the seed condition \eqref{eq:shift} since $a$ is just a spectator. Note that candidates
for new $S_{11}$ that satisfy the first saturation bound $\kk_{11}\le 4$ (at most four $F_p$ factors) are
\be
S^{\rm min}_{11},\ S^{\rm gold}_{11},\ S^{\rm min}_{11}\cdot S^{\rm gold}_{11},\ S^{\rm min}_{13},\ S^{\rm min}_{14},\ S^{\rm min}_{15},\ S^{\rm min}_{16},\ S^{\rm gold}_{12},\ S^{\rm gold}_{13},\ (S^{\rm min}_{11})^2,\ (S^{\rm gold}_{11})^2.
\ee
Only the following four of these satisfy the second saturation bound $\kk_{12}\le 4$ for $S^{\rm new}_{12}$ obtained from $S^{\rm new}_{11}$ with the fusion angle $u^{1}_{12}$
\be
S^{\rm min}_{11},\ S^{\rm gold}_{11},\ S^{\rm min}_{13},\ (S^{\rm min}_{11})^2.
\ee
Furthermore, each of these satisfies $\kk_{ab}\le 2M_{ab}$ for all $a,b=1,\cdots,8$.
These are the minimal, golden, diagonal, and saturated seeds, summarized in TABLE \ref{tab:four}.
\begin{table}[h]
\centering
\begin{tabular}{lccc}
\hline\hline
completion & seed $S_{11}$ & $\mathbb{N}$ & $c_{\rm UV}$\\
\hline
minimal & $[1][11]$ & $\one$ & $21/22$\\
golden & $[7][13]$ & $\AFib^{\oplus4}$ & $25/14$\\
diagonal & $[2][10][12]$ & $\GE$ & $15/2$\\
saturated & $([1][11])^2$ & $2\one$ & $31/2$\\
\hline\hline
\end{tabular}
\caption{The four UV-complete seeds selected by the saturation criterion;
$[p]\equiv F_p$ as in Table~\ref{tab:RL}.}
\label{tab:four}
\end{table}

\section{Inversion relations and R-L adjacency matrices}
\label{sec:SMfolds}
Using the Chebyshev identity \(2T_1(x)T_n(x)=T_{n+1}(x)+T_{|n-1|}(x)\) on \eqref{eq:e8inversion}, we can rewrite it as
\be
2 T_{15}(x)n(1)+2\sum_{p=1}^{15}\left[(1+\delta_{p,14})n(p+1)+n(p-1)-\GE n(p)\right]T_{15-p}(x),
\ee
where the second sum vanishes due to the relations satisfied by the exponent matrices
\be
(1+\delta_{p,14})n(p+1)+n(p-1)=\GE n(p).
\ee
This leads to
\be
\big[(q+q^{-1})\one-\GE\big] \Psi(q)
=(q^{15}+q^{-15})\mathbb{N},\qquad n(1)=\mathbb{N}.
\ee

\section{Numerical TBA/$Y$-systems}

From the inversion relations, we can derive the universal TBA equations as follows:
\be
\epsilon_a^R(\theta)=r\nu_{a}^R+\varphi\star\sum_{b=1}^{8}
\left[(\GE)_{ab}\,\left(\mathcal{L}_b^R-\nu_b^R\right)-\mathbb{N}_{ab}\,L_b^L\right](\theta),
\qquad a=1,\dots,8,
\ee
and another set of equations by exchanging $R\leftrightarrow L$. 
We use the standard convolution $\star$ and $\nu_{a}^R=m_a\,e^{\theta}, \nu_{a}^L=m_a\,e^{-\theta}$. The universal kernel $\varphi$ and $L$-functions are defined by
\be
\varphi=\frac{1}{2\cosh\theta},\qquad \mathcal{L}_a^{R/L}=\log\!\left[1+e^{\epsilon_a^{R/L}(\theta)}\right],\qquad
L_a^{R/L}(\theta)=\log\!\left[1+e^{-\epsilon_a^{R/L}(\theta)}\right],\qquad
\epsilon_a^L(\theta)=\epsilon_a^R(-\theta).
\ee
In terms of these, the scaling energy of the ground state and the effective central charge are given by
\be
E(r)=\frac{\pi c_{\rm eff}(r)}{6r}=-\frac{1}{2\pi}\sum_{a=1}^{8}\int_{-\infty}^{\infty}d\theta\left[\nu^R_a(\theta) L_a^R(\theta)+\nu^L_a(\theta) L_a^L(\theta)\right].
\ee

\begin{figure}[h]
\centering
\includegraphics[width=1\textwidth]{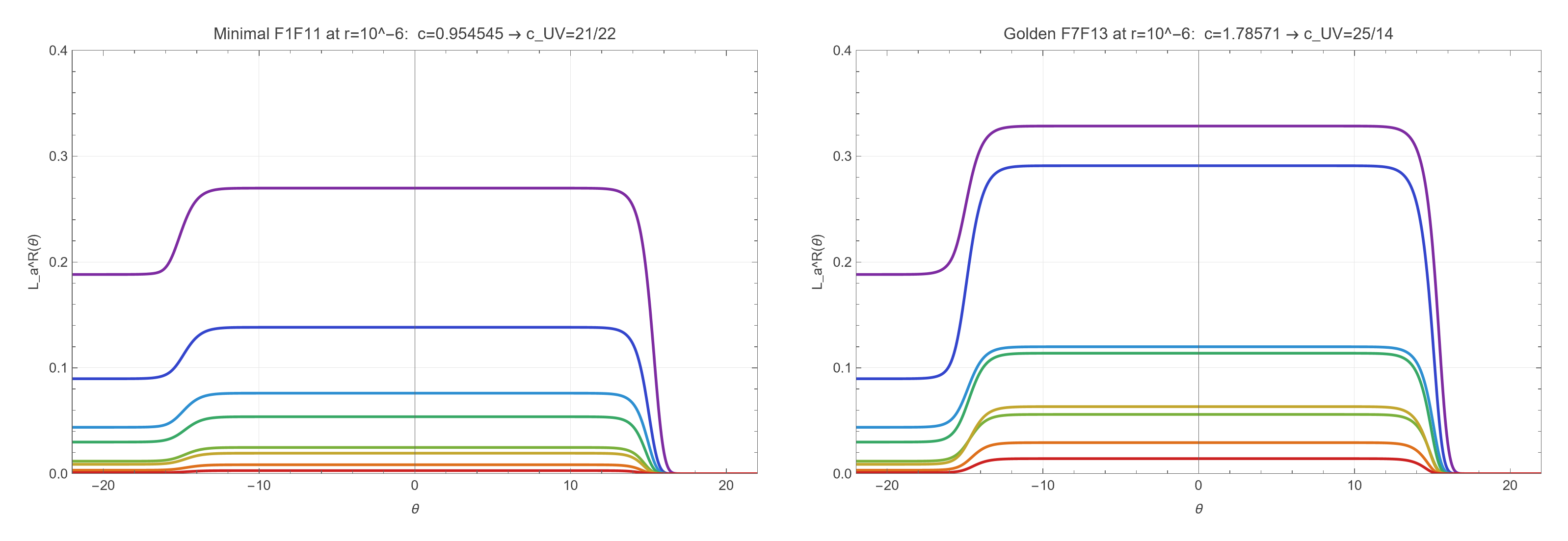}
\caption{$L$-functions $L_a^{R}(\theta)$ for minimal and golden completions at
$r=10^{-6}$.
In each panel the eight curves are ordered by mass and colored from purple
(lightest, top plateau) to red (heaviest, bottom):
$L^R_1\,(m_1{=}1.000)$,
$L^R_2\,(m_2{=}1.618)$,
$L^R_3\,(m_3{=}1.989)$,
$L^R_4\,(m_4{=}2.405)$,
$L^R_5\,(m_5{=}2.956)$,
$L^R_6\,(m_6{=}3.218)$,
$L^R_7\,(m_7{=}3.891)$,
$L^R_8\,(m_8{=}4.783)$,
in units $m_1=1$.
The wide flat plateau is the UV conformal fixed point, where the dilogarithm
sum of the plateau values reproduces $c_{\mathrm{UV}}$; the curves drop to the
trivial vacuum for $\theta\gtrsim\log(2/r)$.}
\label{fig:Lfun}
\end{figure}

A standard iteration algorithm for the TBA equations is applied to generate the RG flows as a function of the scale $r$, as shown in FIG.\ref{fig:ceff}.
We can also obtain the plateau values by plotting the $L$-functions in the deep UV as shown in FIG.\ref{fig:Lfun}. 
For the golden solution, the plateau values of $Y_a^R$ are given by
\be
Y_a^R\simeq(2.5717,2.9608,7.8475,8.2959,15.287, 17.370,33.589,69.871),
\ee
giving $c_{\rm UV}=25/14$
in terms of the Rogers dilogarithm sum. 
The $Y$-system can also be used to find the half-period, which is related to the dimension of the perturbing operator. For the golden case,
$P_{\rm TBA}=35$, which corresponds to
$\Delta_{\rm rel}=2/7$. 

\section{Affine-branching spectrum of the diagonal $su(2)_5$ coset}
\label{sec:SMbranching}

The diagonal $su(2)_5$ coset of Eq.~\eqref{eq:ccoset} has primaries
$(\ell_1,\ell_2;\ell_3)$ with 
\be
0\le\ell_1,\ell_2\le 5, \quad 0\le\ell_3\le 10,\qquad
\ell_1+\ell_2+\ell_3\in 2\mathbb{Z},
\ee
modulo the field
identification~\eqref{eq:fieldid}.  

The physical conformal weight is computed from affine branching functions, defined by
characters $\widehat\chi^{(k)}_{\ell}$ of the $su(2)_k$ WZW model as follows:
\be
\widehat\chi^{(5)}_{\ell_1}(q,z)\widehat\chi^{(5)}_{\ell_2}(q,z)
 =\sum_{\ell_3=0}^{10} b^{\ell_3}_{\ell_1\ell_2}(q)
  \widehat\chi^{(10)}_{\ell_3}(q,z).
\ee
From this, the holomorphic dimensions of the primary states of the coset CFT are given by
\be
h_{\rm phys}=h^{(5)}_{\ell_1}+h^{(5)}_{\ell_2}-h^{(10)}_{\ell_3}+n_{\min},
\qquad
h^{(k)}_\ell=\frac{\ell(\ell+2)}{4(k+2)},
\ee
where $n_{\min}$ should be chosen to make the dimensions non-negative.

From this formula, we obtain low-lying dimensions as shown in Table~\ref{tab:genuineSpec}.
The $h=1/7$ entry of Table~\ref{tab:genuineSpec} is found at
\be
(1,3;4)\sim(3,1;4)\sim(2,4;6)\sim(4,2;6)
\ee
as given in Eq.~\eqref{eq:phi-orbit-doubled} of the main text.
\begin{table}[h]
\centering
\begin{tabular}{ccc||ccc}
\hline\hline
$h_{\rm phys}$ & decimal & $(\ell_1,\ell_2;\ell_3)$ &
$h_{\rm phys}$ & decimal & $(\ell_1,\ell_2;\ell_3)$\\
\hline
$0$       & $0.00000$ & $(0,0;0)$ & $25/112$ & $0.22321$ & $(0,3;3)$\\
$5/112$   & $0.04464$ & $(0,1;1)$ & $79/336$ & $0.23512$ & $(1,4;5)$\\
$1/21$    & $0.04762$ & $(1,1;2)$ & $37/112$ & $0.33036$ & $(1,2;1)$\\
$1/14$    & $0.07143$ & $(2,2;4)$ & $5/14$   & $0.35714$ & $(0,4;4)$\\
$9/112$   & $0.08036$ & $(1,2;3)$ & $17/42$  & $0.40476$ & $(2,2;2)$\\
$31/336$  & $0.09226$ & $(2,3;5)$ & $10/21$  & $0.47619$ & $(1,3;2)$\\
$5/42$    & $0.11905$ & $(0,2;2)$ & $57/112$ & $0.50893$ & $(2,3;3)$\\
$1/7$     & $0.14286$ & $(1,3;4)$ & $25/48$  & $0.52083$ & $(0,5;5)$\\
$3/14$    & $0.21429$ & $(1,1;0)$ & $4/7$    & $0.57143$ & $(2,2;0)$\\
\hline\hline
\end{tabular}
\caption{Low-lying genuine conformal weights of the golden coset.  The
$h=1/7$ entry is the perturbing orbit selected by the $Y$-system.}
\label{tab:genuineSpec}
\end{table}

\end{document}